\documentclass[9pt]{article}
\usepackage{epsfig}
\def\ArtWork#1{\noindent\hfill\epsfbox{#1}\hfill}

\begin{document}
\setlength{\parskip}{0.5pc}
\setlength{\parindent}{0.3cm}
\addtolength{\oddsidemargin}{-2cm}
\addtolength{\topmargin}{-2.5cm}

\title{\textbf{On the issues of building Information Warehouses}\thanks{Final version at ACM Comptute 2010, January 22-23, 2010, Bangalore, India.}}
\author{Arijit Laha\thanks{Arijit Laha, SETLabs, Infosys Technologies Ltd., Manikonda, Hyderabad 500019, India., Email: arijit\_laha@infosys.com}}

%\begin{document}
\date{}

\maketitle

\begin{abstract}
While performing knowledge-intensive tasks of professional nature, the knowledge workers need to access and process large volume of information. Apart from the quantity, they also require that the information received is of high quality in terms of authenticity and details. This, in turn, requires that the information delivered should also include argumentative support, exhibiting the reasoning process behind their development and provenance to indicate their lineage. In conventional document-centric practices for information management, such details are difficult to capture, represent/archive and retrieve/deliver. To achieve such capability we need to re-think some core issues of information management from the above requirements perspective. In this paper we develop a framework for comprehensive representation of information in archive, capturing informational contents along with their context. We shall call it the ``Information Warehouse (IW)" framework of information archival. The IW is a significant yet technologically realizable conceptual advancement which can support efficiently some interesting classes of applications which can be very useful to the knowledge workers.
\end{abstract}


\section{Introduction}
Knowledge works are typically aimed at solving ``wicked problems" \cite{Kuntz:70}. They are hugely complex and dynamic interplay of human intellect and environment involving reflective thinking \cite{Dewey:97}. The informational requirements of the people engaged in them are also of extremely complex and evolving in nature  \cite{Conklin:96,Markus:02}. Markus et al. \cite{Markus:02} categorized such works, e.g., basic research, new product development, strategic business planning, organization design etc., as ``Emergent Knowledge Processes (EKP)" and studied the problem of building IT-based support systems for them. Their work illustrated the inadequacy of most of the commonly practiced design approaches for such support systems. Such systems are important part of overall Knowledge Management (KM) infrastructures of organizations. However, the behavioral and socio-cultural concerns \cite{Alavi:01} in building typical KMS further aggravate the problem.

In a typical knowledge work, a large volume of information is collected and processed by the knowledge workers. A significant portion of the required information may come from the worker's own experience/expertize and from other workers and experts by means of social interactions. Nevertheless, much of the information come from \emph{archived documents} \cite{Markus:02}. In modern organizations, such archives and the means  to access them are by and large based on computer/IT systems. Use of such systems, from technological perspective, usher us from the era of ``paper documents" to that of ``electronic" or ``digital" documents. However, we are yet to exploit the full potential of that transition. Our practices regarding creation, archival and retrieval of information, despite the replacement of \emph{paper documents} by the \emph{electronic documents} as the containers of information, is still influenced heavily by past several centuries of paper based information management practices.

For example, for a knowledge worker to perform a task, he/she needs to access and study information related to past instances of similar or related tasks. Ideally, the worker would like to get information on not only \emph{what} has been done, but also \emph{why} and \emph{how} they have been done as well as what were the results or impacts of doing so. To answer such questions, information need to be available about argumentative supports, i.e., the reasoning processes followed, in the \emph{arguments made, decisions taken, alternatives selected etc.} in various steps of the past task instance. This requirement can be satisfied only if detailed informational contents along with their contexts at proper granularity levels are captured, archived and delivered to the users. Unfortunately, our conventional document-centric information management strategies are seldom adequate for the purpose \cite{Conklin:96}.

In this paper we shall explore some possibilities provided by the electronic documents in order to address the informational needs of the knowledge workers. The problem, as we shall see, will require examining a number concepts from several areas of study. Further, there is multiple facets of the problem, including capture, archival, exploration and delivery of informational contents along with their contexts. Addressing all of them adequately is beyond the scope of a single paper. So, in this paper, we shall concentrate mainly on the issues related to archival. In the process, we  shall develop a archival framework for granular informational contents and their contexts, which we shall call the ``Information Warehouse" or IW. While IW can be used to support various sophisticated applications, we shall briefly illustrate its efficacy in context of supporting a new class of information systems, the ``Knowledge Work Support Systems" or KWSS, a term used by several researchers, including Burstein and Linger \cite{Burstein:03} in context of their Task-based Knowledge  Management (TbKM) for studying and analyzing the problem of building task-specific information management systems.

The rest of the paper is organized as follows: in section \ref{need} we illustrate with a real-life scenario the need for IW-like systems and examine some issues typically preventing us to conceive them; issues regarding systematic and comprehensive organization of informational contents along with their relevant contexts are explored and resolved in section \ref{organization}; section \ref{usage} demonstrates the efficacy of IW in building advanced and sophisticated applications especially the KWSS; we conclude the paper in section \ref{conc}.

\section{Need for Information Warehouse (IW)}\label{need}

In the space of \emph{structured data}, the concept of Data Warehouses (DW) is developed in order to provide archival environments where \emph{integrated}, \emph{quality-controlled} and \emph{re-organized} historical data can be used easily for complex analysis. These issues are equally, if not more, important in the space of \emph{unstructured textual information}, or simply \emph{information}, used extensively in knowledge works. However, no such concept exists in the information space. The Information Warehouse (IW) framework is an attempt to provide comprehensive means to understanding and addressing similar issues in context of unstructured, largely textual information.

\subsection{An example to illustrate the need for IW}
Coy et al. \cite{Coy:98} in their study on \emph{Hedge Fund crisis} of 1998 identified three computer-system related issues,
\begin{enumerate}
  \item the assumptions that were embedded in the computer models were invalid,
  \item there was a breakdown in the historical patterns on which the assumptions were based and
  \item the computer models were ``black box" models that were allowed to make automated decisions without intervention of human experts.
\end{enumerate}
Sound familiar? Yes, similar points are being made in context of the much graver \emph{global recession} we are going through. We should have learnt enough from such past debacles. Actually, from computer systems perspective, there are some deep-rooted difficulties in \emph{really} learning and \emph{more importantly} applying the lessons learnt to avoid such happenings.

Such systems typically fall under the category of Decision-Support Systems (DSS) \cite{Keen:78}. The problem of building DSS has been extensively studied during 1980s and 1990s \cite{Power:03}. Many sophisticated and complex DSS found their way in assisting people in complex operations across many businesses and other organizational domains. In many domains, e.g., manufacturing processes, where the operational environment is stable and well-controlled, they have been found extremely successful. In other domains, especially socio-economic ones, characterized by continuous and often unpredictably evolving environment, their use might suffer from a significant problem. Given the scale and complexity of operations typically undertaken by such organizations, use of DSS \cite{Power:03} is possibly unavoidable. However, they need to be aware of the limited span of usability of DSS in face of the changes in environment and make sure that the DSS adapts \cite{Vicente:00} to the changes. This, in turn, requires that the DSS be routinely evaluated and updated, i.e., maintained up-to-date.

The task of effectively maintaining complex DSS is, at least in practice, extremely difficult one. Typically, building a DSS is a complicated undertaking requiring participation of a good number of experts, e.g., domain experts, technology experts, knowledge engineers, developers etc. Use of a lot of \emph{specialized knowledge} takes place in the process, which culminates in development of artifacts such as algorithms, rule-bases, system of equations, neural networks etc. which drive the DSS. Naturally, much of these appear quite esoteric to the intended user community and the system is treated as a \emph{black box} \cite{Coy:98} by them. Down the line, when the time comes to review and update the system, a team of experts (even if it include the original ones, which seldom happens in practice) need to know not only \emph{what} the DSS and its various parts do, but also the \emph{how} and \emph{why} do they do what they do. Typically, information at such level of details are not available.

One may argue that it takes only \emph{greater effort} in documentation to capture the details. However, in practice, for such complex systems, it is extremely difficult to do so using usual means of \emph{linear} document-oriented information management \cite{Conklin:96}. We need to explore some innovative ways to help people in capturing, archiving and retrieving detailed information. \emph{Note} that, the problem described here is \emph{representative of a large number of problems}, including other development/maintenance tasks for complex software, other complex systems and in general, a large gamut of \emph{knowledge-intensive tasks with large informational requirements}. The EKPs (basic research, new product development, strategic business planning, organization design etc.) mentioned by Markus et al. \cite{Markus:02} are prime examples of them.

\subsection{Documents as containers of Information}\label{sub:info-doc}
For several centuries  human civilization has practiced \emph{paper document} based information management. Among the essential characteristics of a paper document are (1) it is a physical, 2-dimensional object, (2) it contains information encoded in form of visible inscriptions on its surface and (3) once it is created, the organization of informational contents, reflecting their logical (comprised of the elements like chapter, section, subsection etc) and argumentative (interrelationships among contents demonstrating reasoning process)  structures remains unchanged, i.e., it presents same view of its contents to every user \cite{Buckland:97,Lavy:94,Pedauque:03}.

Introduction of computers and thus \emph{electronic documents} opened up a wide range of possibilities which we are yet to understand fully and exploit for our benefit \cite{Pedauque:03}. Invention of Hypertext \cite{Conklin:87,Trurana:07} constitutes a great step towards realizing many of these opportunities. However, the flexibility introduced by hypertext documents is largely limited to the logical aspects of the content organization \cite{Millard:05}. Thus, creator of the documents still need to \emph{linearize} the non-linear \emph{contextual structures} of the contents \cite{Lavy:94}, leaving to the potential readers the task of reconstructing them to the best of their ability. In this paper we contend that it is possible to \emph{enhance} the ability of the users (creators as well as readers) in dealing with information with appropriate but somewhat innovative use of available technologies.

Technologically, most crucial difference stems from the fact that unlike a paper document, its electronic counterpart does not have any humanly discernible physical reality. It is just a string of digital bits and to make use of it we must use a computer and a software which is programmed to process those bits. So, one may ask, which one is \emph{the} document, the collection of bits or what is visible on the computer screen due to logical processing of the bits by the software? Thus, the existence of the electronic document, is more of a \emph{logical} matter than \emph{physical}. Extending this line of arguments Buckland \cite{Buckland:98} observed
\begin{quote}
    ... whatever is displayed on the screen or printed out is a document. One might say that the algorithm is functioning as a document, as a dynamic kind of document, ... it would be consistent with the trend, ... towards a defining document in terms of function rather than physical format.
\end{quote}

This opens up many exciting possibilities regarding presenting different \emph{customized} views of a document to different interested parties (e.g., creators and different readers with different motivations). P\'{e}dauque \cite{Pedauque:03} describes the possibility as one where a document in its archived form is a sort of \emph{puzzle} whose pieces can be \emph{assembled on request} by the machine. Thus, an electronic document can be viewed as a modern \emph{magic slate} on which various informational items (text as well as anything else those can be digitized) according to readers' \emph{preferences and needs} \cite{Pedauque:03}, and thus can convey greater significance.

With evolving maturity in using electronic documents as means of information management, an emerging trend can be observed. The trend is characterized by information contents with \emph{smaller granularity}, \emph{richer linkage} among contents and various types of \emph{structuring} (logical: HTML/ XML, semantic/metadata: RDF/OWL etc.) around the contents \cite{Volkel:07}. These, in general, lead to much better utilization of informational resources. Extending this trend, especially with focus on meeting information requirements of knowledge-intensive tasks, V\"{o}lkel \cite{Volkel:07} envisaged the ``Knowledge Models" of documents, defined as
\begin{quote}
    ... a set of items, linked by directed, typed relations. An item is either an information atom or an entity composed of other items. Each item can contain a small piece of text, such as a paragraph, a sentence, a single word, or a reference to an external web or desktop resource. Relation types are also labelled with items.
\end{quote}

While V\"{o}lkel's idea is interesting one, realizing it, especially in the generic manner envisaged by V\"{o}lkel, can be extremely difficult. However, the problem may take somewhat manageable proportion if we restrict ourselves regarding the domain of application. In this paper, we are concentrating on the problem of building \emph{informational support systems for knowledge works of professional nature}. This will allow us to consider issues such as suitable granularity levels to be represented by the information atoms or \emph{informational elements} (IE) and their relationship types, with respect to \emph{a particular task-type in a particular domain} at a time. Even with the identification of the task-type and domain, putting together a whole, consistent and most importantly, technologically viable framework can be a significant challenge.

\section{Organization of information in IW}\label{organization}
The primary goal of the Information Warehouse (IW) framework, is to provide a framework for representing informational contents related to \emph{knowledge works of professional nature}, along with their naturally non-linear contextual support structures in sufficient detail to meet the informational requirements of knowledge workers. The IW framework is an innovative extension of the trend noted above in terms of granularity, linkage and structures of contents. In the following we explore proper usage of these notions in context of knowledge works.

\subsection{Granular contents and their contexts}
In knowledge works, access to informational contents at suitable granularity level is essential for their efficient utilization. But given a knowledge work or task, what level can be \emph{suitable}? Further, we need to link the contexts with the granular contents. The notion of context, in its entirety, is vast and highly complex one \cite{Dourish:04}. It is impossible to capture all of it. So, we need to determine which part of context we need to work with. Finally, since we intend to manage the information by means of IT systems, how do we represent the granular contents and their contexts computer systems can use them \emph{algorithmically}?

\subsubsection{The Activity-level Granularity}
Knowledge workers need information while performing their works. But do they need to have all the information required for the whole task (assuming it is substantially complex/large) at once? The answer is an emphatic \emph{No!}. It is well recognized in computer science \cite{Newell:72} as well as cognitive psychology \cite{Baddeley:74} that the problem-solving takes place in ``working memory" of human brain, which has fairly limited capacity compared to the long-term memory that holds conceptual knowledge. As a result, to perform a task, we need to decompose them in smaller, manageable units of work, the \emph{activities}. Burstein and Linger \cite{Burstein:03} defined a task as \emph{a system of activities} comprised of structure and processes. Johnson and Johnson \cite{Johnson:91} assert that we learn about tasks as well as their perform them following a structure constituted of components such as actions, objects and goal substructures. Similar notion of task-structure forms basis of the Activity Theory \cite{Nardi:96}.

While there is are some differences in the terminologies used, all these theories equivocally assert that \textbf{in a knowledge-intensive task, at a time, a worker is engaged in \emph{one activity} as part of the task}. Consequently, his/her informational usage are also related to that activity. Thus, the \emph{granularity level of accessibility} of the informational contents should be commensurate with those \emph{developed/created} and \emph{consumed} at the \emph{level of activities}.

\subsubsection{Two useful types of contexts}
Activities occur ``within" their contexts \cite{Dourish:04}. Consequently, information related to activities are only useful if their context is known. In other words, to make good use of information, a knowledge worker need to understand various aspects of the context within which the information has come into existence and compare it with the current work context. Unfortunately, in a typical IM environment, the information contained in documents seldom carries enough contexts \cite{Conklin:96}. Thus, the users of the documents has to by and large depend on their knowledge/experience to reconstruct and understand the contexts. However, as Markus et al. \cite{Markus:02} observed, there is, in general, significant variation of knowledge and ability levels among the potential users. This lead to wide variation of the quality of the performance across the user community. In IW we propose to archive along with the granular informational contents, their relevant contexts.

There can be a large gamut of context types \cite{Dourish:04}. Here we are interested in the type of contexts those serve mainly two purposes, (1) allow experienced knowledge workers to efficiently access and utilize archived information and (2) aid learning about the task, so that less knowledgable workers, at least to some extent, can overcome easily their deficiency. From this perspective we recognize two types of contexts, (1) the \emph{categorical context}, representing the general task-type characteristics of the informational elements and (2) the \emph{instance-level} or \emph{episodic context}, representing the relationships among informational contents developed during performance of \emph{task instances}. The former can be captured by links representing relations among various activity and content categories, while the later can be represented by links among the granular contents themselves. Further, associations between contents and their types as represented in the categorical context, facilitates sophisticated analysis of the contents.

\subsection{Representing the granular contents and their contexts}
At conceptual level, the IW can be thought as a framework for elaborate annotation of the contents. Thus a theoretical treatment can be pursued in a line similar to that proposed by Agosti and Ferro \cite{Agosti:07}. However, in this paper we follow a functional approach in our treatment of the problem. Clearly, in order to realize the the IW framework by means of IT systems, we need to incorporate in it well-defined and algorithmically manipulable structural elements representing the granular contents and their contexts. In the following we develop an approach towards achieving it.

\subsubsection{The Structures}
In IW, the central concept is that of an \emph{activity-based decomposition} of tasks and corresponding information granules. Based on the concept of the task-structures, for a given task type, we can model the typical pattern of performance followed during its execution. Let us call it the ``Activity-Flow" model of tasks. The activity-flow of a task provides a representation of the task in terms of the constituent activities and their various relationships. Evidently, this is an instance-independent \emph{categorical model} of the tasks where the activities are categories, rather than instances of performance. The activity-flow model forms the mainstay of the categorical context for a task. Given an activity category, it is easy to determine type of granular information developed by its performance as well as the typical reasoning process used and other content types used in the process. In other words, with an activity type identified in the action flow, we can link with it content types and their interrelations typically associated with the activity type. Further, for each of the identified types (activities and contents) we can link them with suitable domain-semantics. These together provide a robust and easily comprehensible representation of the categorical context.

The goal of the episodic context is to capture the argumentative supports and provenance of the contents. To capture them we shall consider two possible scenarios. Firstly, one content might have its genesis in order to satisfy, (may be partially) the need of developing another content. We shall call it a \emph{demand-satisfaction} (DS) relationship. The other one is that of reference, which we shall denote as the \emph{referential support} or RS relationship, where a content may not be created explicitly to fulfill the need of other, but provide referential support for the development of the later. Given an activity-flow model of a task, for each of the activity categories, it is easy to analyze the details informational requirements, processing patterns, creation of contents and their support. Such analysis allows us to identify for each activity category,  typical content categories related to it and many of the  DS and RS relations among them. This can be again made part of the categorical context describing the typical content organization that occurs during performance of task instances. The contextualization of contents can be further enriched by accommodating the domain-semantics of the activity and content categories in the framework.

To summarize the above, for a particular granular content or informational element (IE) and corresponding activity, the episodic context allow us to enquire what, how and why \emph{happened} before and after \emph{the activity} while the categorical context allow us to find what , how and why \emph{happens} before and after \emph{such activity}. Clearly, the organization of information, i.e., \emph{contents and their contexts}, in IW has the characteristic of a number of interrelated ``graphs" or ``networks". For a task-type, its activity-flow forms a graph with nodes representing activities and the edges signifying their various relations. Similarly, the content/IE categories also form a graph with each category as a node. The domain-semantics of the categories and their relationships also can be represented by a graph. The ``task instance (TI)" information is represented as a graph with contents or IE instances as nodes and the DS and RS links as edges. The issues related to modeling and representing information in the networked form discussed above is studied extensively under the ``graph data modeling" techniques \cite{Angles:08}.

\subsubsection{An illustration}

\begin{figure*}%1.0
\epsfxsize=0.8\hsize \ArtWork{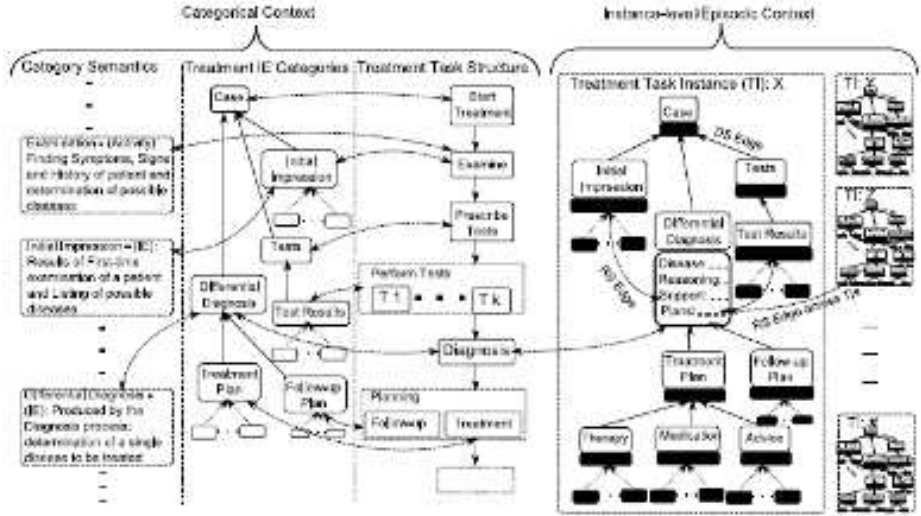} \caption{The categorical and episodic contexts.}\label{cont}
\end{figure*}

The network structure of the IW contents is illustrated graphically in figure \ref{cont} with the help of a simplified and high-level activity-flow for the task-type of patient-care. Here a task instance is the ``treatment of a patient" by a physician. Let us consider the granular content ``differential diagnostic" or DD for brevity, as shown in the figure. The episodic context of DD is formed by other contents in the case which include the top-level task description in content ``Case", as well as various plans related to treatment of the identified disease. To perform the diagnosis, the doctor \emph{refers} to the findings of the examination of the patient which is captured in the content ``Initial Impression" as well as results of various diagnostic tests conducted. He/she may also refer to past cases across TI. The categorical contexts of the content DD is provided by the treatment task structure by associating it with the activity/process ``Diagnosis". In the content categories, in turn, the typical informational context of DD is described. Further, the category semantics provide readily available interpretation of both the activity categories and the content categories. For example, here DD as an content type represents a specific technical meaning from the healthcare domain.

\section{Using the IW}\label{usage}
An archive built following the IW framework can be used in various ways. It gives us a
systematic and comprehensive means to analyze and design archival schemes for informational contents at flexible granularity levels along with their contexts. An IW design is conceptual one which can be converted into suitable logical design based on the particular technology, e.g., RDBMS, XML database, Object database etc., used to implement the archive. Such an archive, once implemented, provides a robust infrastructure for building many powerful applications. The related issues, such as populating the archive, exploring the contents, various ways of delivering the contents for efficient use etc. are to a large extent depends on the particular application to be built on top of the archive. Here we shall illustrate some ways to address those issues in context of a novel class of information systems, the Knowledge Work Support Systems or the KWSS.

\subsection{The Knowledge Work Support Systems}
While a few researchers used the term KWSS, it is the works of Burstein and her co-workers on Task-based Knowledge Management or TbKM \cite{Burstein:03}, that
provide clarity in their characteristics. They developed the TbKM framework in order to understand the nature of the informational supports needed by knowledge workers on the basis of the specific types of knowledge-intensive tasks they perform. They conceptualized a KWSS as an IT-based support system which is aware of the structures and processes of the task and can provide targeted supports to them. While TbKM successfully underline various characteristics desirable in KWSS, from technology perspective, they do not consider the possibility of one integrated infrastructure for systematically supporting all the components or modules of a KWSS. While they suggest using various ``Intelligent DSS" in more focused manner during the activities, this approach fails to address the problem faced by knowledge workers in typical modern organizations, as pointed out by Markus et al. \cite{Markus:02} that there is rather a ``glut" of tools which are used without proper integration to a cohesive work process, and thus often becomes source of additional problems.

\begin{figure}%1.0
\epsfxsize=0.5\hsize \ArtWork{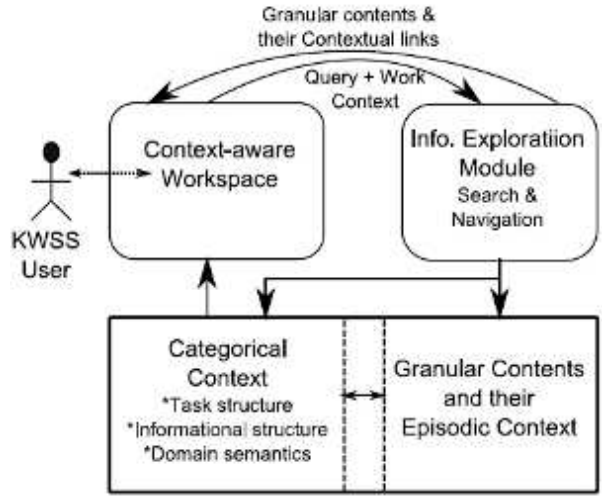} \caption{The basic structure of a typical KWSS.}\label{kwss}
\end{figure}

Here we envisage a KWSS, built on the top of an IW archive, can deliver the functionalities envisaged in TbKM \cite{Burstein:03} and do so in such a manner that addresses the concerns raised in \cite{Markus:02}. While the very basis of IW is the nature of informational requirements of typical knowledge workers, in the following we shall briefly outline how the knowledge workers can be enabled to easily populate the archive and retrieve information from it while using a KWSS. The basic structure of a typical KWSS is depicted graphically in figure \ref{kwss}.

\subsubsection{Populating the IW}
Once we target a specific task-type to build a KWSS for, it is not difficult to study and encode its task-structure and other elements of the categorical context, as defined above, for the targeted task \cite{Burstein:03,Dourish:04,Nardi:96}. In a KWSS, the predefined categorical context of the target task-type plays a central role of the \emph{definition of the task} and used extensively in multiple ways by its various components. Actually, in building a KWSS, the first step is to design, encode and deploy it. Once it is deployed and ready to be accessed by various components of the KWSS, the KWSS is ready for use.

\begin{figure}%1.0
\epsfxsize=0.5\hsize \ArtWork{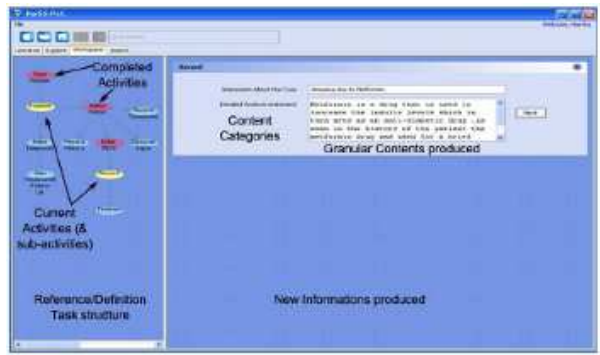} \caption{The context-aware workspace for a prototype KWSS.}\label{workspace}
\end{figure}

Performance of knowledge works lead to production of information. Let us see how \emph{new information} produced while a user is performing an instance of supported task-type can be captured and archived. With the availability of the definitional elements (i.e., the categorical context) of the task, a context-aware workspace \cite{Winograd:01} can easily track the progress of the task and maintain the work context. Therefore, whenever a user, working through the workspace, develop new informational contents, they are automatically identified with their proper context by the workspace. Thus, the informational contents and their context can be archived without additional effort by the user. This ease of recording information also address a crucial behavioral impediment as observed by Conklin \cite{Conklin:96} related to the extra effort needed by workers to produce separate documentations. A sample screenshot of the workspace in patient-care KWSS is shown in figure \ref{workspace}.

\subsubsection{Information Retrieval from the IW}
The core concepts and algorithms of Information Retrieval (IR) do not presuppose any particular type of entities as the containers of information \cite{Faloutsos:95}. They work with mathematical (or rather geometrical) vector-space representation of the information. Thus, conceptually, the IR techniques used for retrieving conventional documents are equally useful for the granular contents. Only some minor systemic modifications in their implementations need to be made to account for difference in types of information containers. Actually we envisage much more enhanced IR capabilities with use of the elements of categorical context as bases of building indexes.

\begin{figure*}%1.0
\epsfxsize=0.8\hsize \ArtWork{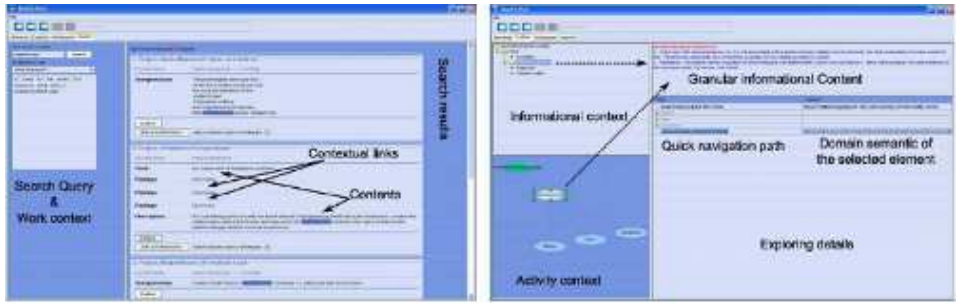} \caption{The search and exploration (navigation) interfaces for a prototype KWSS.}\label{search}
\end{figure*}

While the user is working through the context-aware workspace, the search functionalities can be invoked from within the workspace. This allows us to easily export, along with the query, detailed work-context of the query to the search module of the KWSS. This makes possible very focused use of the IR facilities in providing relevant information. Further, along with the easily consumable granular contents, IR also returns their contextual links to other contents, categories and semantics. These links allow the user to navigate for getting a clearer understanding of the contents. Both of these search and navigation functionalities together provide a very rich and easy-to-use environment for information retrieval from IW and their efficient utilization. Further, due to the well defined association of the contents with their semantic contexts, the IR system can easily extended to incorporate semantic search \cite{Mangold:07} capabilities. A possible design of user interfaces for search and navigation is illustrated in figure \ref{search} through screenshots from a patient-care KWSS prototype.

In context of KWSS we can think of many more sophisticated functionalities added to it which can be built exploiting the IW. Such possibilities include recommendation \cite{Adomavicius:05}, collaboration \cite{Rama:06}, argumentation \cite{Conklin:88} etc. Availability of the detailed contents and their contexts can make significant difference in capabilities of them.

\subsection{Some other applications}
Apart from KWSS, the IW can support many more applications in innovative and/or enhanced manners. One of the problem recently attracting a lot of attention is that of ``Expertise Location" \cite{McDonald:98} in big organizations. It is easy to see that the analysis of the IW contents can build vary good evidence-backed profiles of experts whose work instances are recorded in IW. In the space of Organizational Learning, one of the most difficult problems is that of ``double-loop learning" \cite{Argyris:78}. Again, given the details of information captured in IW in an organized manner, it can be easily systematically analyzed to unearth subtle causal relations among the activity and task instances. Such analysis can help immensely in double-loop learning. On of the most exciting possibility we are currently exploring is that of using an IW as an platform for building a number of KWSSes and other applications on top of it so that significant portion of organizational KM requirements can be addressed through it.

\section{Conclusion}\label{conc}
In this paper we have pointed out the need for detailed but at suitably granular level informational contents along with their contexts in performing knowledge-intensive tasks of knowledge workers. We have shown that some re-thinking in the way the notion of electronic documents are used can address the issue substantially. We have developed a framework, the Information Warehouse or IW, for systematic archival of granular informational contents and some of their crucial contexts. We have also outlined how the IW can be used in KWSS and other sophisticated applications. We believe that IW is a plausible and technologically viable way of managing information for use in tasks where reflective thinking by human actors is a necessity. We are currently engaged in exploring various frameworks, methodologies and technology stacks for building IW and KWSS.

\bibliographystyle{plain}
\bibliography{ckim}

\end{document}